**Graphene formed on SiC under various environments:**
**Comparison of Si-face and C-face**

N. Srivastava, Guowei He, Luxmi, P. C. Mende, and R. M. Feenstra
Dept. Physics, Carnegie Mellon University, Pittsburgh, PA 15213

Yugang Sun
Center for Nanoscale Materials, Argonne National Laboratory, Argonne, IL 60439

**Abstract**
   The morphology of graphene on SiC {0001} surfaces formed in various environments including ultra-high vacuum, 1 atm of argon, and $10^{-6}$ to $10^{-4}$ Torr of disilane is studied by atomic force microscopy, low-energy electron microscopy, and Raman spectroscopy. The graphene is formed by heating the surface to 1100 – 1600°C, which causes preferential sublimation of the Si atoms. The argon atmosphere or the background of disilane decreases the sublimation rate so that a higher graphitization temperature is required, thus improving the morphology of the films. For the (0001) surface, large areas of monolayer-thick graphene are formed in this way, with the size of these areas depending on the miscut of the sample. Results on the ($000\bar{1}$) surface are more complex. This surface graphitizes at a lower temperature than for the (0001) surface and consequently the growth is more three-dimensional. In an atmosphere of argon the morphology becomes even worse, with the surface displaying markedly inhomogeneous nucleation, an effect attributed to unintentional oxidation of the surface during graphitization. Use of a disilane environment for the ($000\bar{1}$) surface is found to produce improved morphology, with relatively large areas of monolayer-thick graphene.

**I. INTRODUCTION**
Epitaxial graphene on SiC{0001} has been intensively studied over the past five years as a potential means of producing *large area* graphene for electronic applications.[1] There are two inequivalent faces of SiC{0001} – the (0001) face, also known as the Si-face, and the ($000\bar{1}$) face or C-face. On both surfaces graphene can be formed by heating the SiC in vacuum, causing preferential sublimation of the Si atoms thereby leaving behind excess C atoms which self-assemble into the graphene. Preparation is also possible in other environments including argon,[2,3,4] or disilane,[5] both of which have been shown to improve the structural quality of the film, particularly for graphene formed on the Si-face of SiC.

   Despite this success at forming epitaxial graphene on SiC, a number of unresolved issues remain concerning formation of uniform films consisting of single monolayer (ML) or few ML graphene, particularly on the C-face. It has been demonstrated that graphene formation in vacuum is intrinsically different for the C-face compared to the Si-face, with the growth proceeding in a nearly layer-by-layer manner for the Si-face (for ≳2 ML graphene coverage) whereas on the C-face it proceeds in a more three-dimensional (3D) manner.[6, 7] This difference in the growth modes for the two faces is surely influenced by the different temperatures used in the two cases in vacuum, about 1150°C for the C-face but 1300°C for the Si-face. Another contributing factor is the different interface structures for the two faces: a 6√3×6√3-R30° interface layer forms between Si-face SiC and the graphene,[8, 9, 10, 11, 12, 13] acting as a template for the formation of the graphene,[13] whereas on the C-face there are 2×2 and 3×3 structures that variously occur [13, 14, 15] but do not seem to act as templates. (This difference in interface structures likely affects the formation temperatures themselves).

For graphene formation in vacuum, the formation temperature is determined by the preferential sublimation rate for Si as compared to C from the surface. It has been demonstrated, however, that independent control over the temperature and Si sublimation can be achieved, either by performing the heating in an inert atmosphere such as argon or by using a Si-containing environment such as disilane.[2,3,5] As mentioned above, these techniques have demonstrated improvement in the quality of graphene films for the Si-face of SiC, but for the C-face a similar level of improvement has not been obtained. On the contrary, for heating the C-face in argon[16] it is found that relatively thick islands of multi-layer graphene form, and the reason for this phenomenon was demonstrated in our prior work to be unintentional oxidation of the C-face surface in the argon environment.[7] (This oxidation is found not be an issue on the Si-face, apparently due to the relative stability of that surface under the inert gas environment). For heating in disilane, no prior studies have been performed to our knowledge on the C-face.

Our work on the morphological evolution of graphene on the Si- and C-face of SiC under various environments has been previously presented in a number of papers.[4,6,7,17,18] We briefly describe that prior work here, and we present additional, new data for each of the particular surfaces and environments considered. We discuss in particular how the graphene formed by our preparation method compares with that prepared by other groups using different preparation techniques. This discussion of the preparation-dependence of the graphene provides, we believe, useful insight into the mechanisms for the graphene formation.

## II. EXPERIMENT

Our experiments are performed on nominally on-axis (unintentional miscut ≤0.2°), *n*-type 6H-SiC or semi-insulating 4H-SiC wafers purchased from Cree Corp, with no apparent differences between results for the two types of wafers. The wafers are normally 2 or 3 inch in diameter, mechanically polished on both sides and epi-ready on either the (0001) surface or the ($000\bar{1}$) surface. These wafers are cut into 1×1 cm² samples and the samples are chemically cleaned in acetone and methanol before putting them into our custom built preparation chamber which uses a graphite strip heater for heating the samples.[17] The ultimate base pressure of the chamber (after a thorough bake) is $1\times10^{-10}$ Torr, although under normal operating conditions, employing a load-lock for sample introduction, it is used at a base pressure of about $5\times10^{-9}$ Torr. Prior to graphene formation the samples are generally etched in a 10 lpm flow of pure hydrogen for 3 min at a temperature of 1600°C. This H-etching removes the polishing scratches which arise during the mechanical polishing of the wafers, resulting in an ordered step-terrace arrangement on the surface which is suitable for graphene formation.[19] For the studies in a disilane environment reported below, the wafers were cleaned in disilane rather than by H-etching, by heating to 850°C in $5\times10^{-5}$ Torr of disilane for 5 min, since this procedure is more straightforward than the H-etching.

As described above, the H-etching treatment generally results in well-ordered step-terrace arrays on the surface, with full-unit-cell high steps (1.0 nm for 4H or 1.5 nm for 6H material) are formed on the Si-face and half-unit-cell high steps on the C-face. However, we do sometimes observe less well ordered step arrays following H-etching. This can occur for either the Si-face or the C-face, but is more of a problem for the latter. For the C-face, surface that do *not* form a regular step-terrace array also tend to display a significant number of etch pits on the surface (density of $10^3 - 10^4$ cm$^{-2}$) after the H-etching. Figure 1 compares the typical morphology of a C-face sample displaying few etch pits after H-etching [Fig. 1(a)] compared to one with many etch pits after H-etching [Fig. 1(b)]. We cannot at present say what aspect of the sample or surface produces a regular step-terrace array, or not, but the recent observation of Robinson et al.[20] that a slight miscut (>0.2°) leads to a more regular step arrangement is consistent with our own experience. (It is also possible that a greater number of dislocations on certain wafers might lead to a higher number of etch pits, but we have not to date independently measured these two variables for a range of samples). Importantly, surfaces that do *not* display a regular step-terrace array following H-etching *are* found to have a tendency to form significant amounts of nanocrystalline graphite (NCG) on the surface following the graphitization procedure.[6] This NCG is found to form on both the Si-face and the C-face, but with much greater abundance for the latter.

Before graphitization, the hydrogen is pumped away from the chamber until a desired pressure of $10^{-8}$ Torr is reached. The samples are then either heated in this vacuum for 10 to 40 min at temperatures ranging from 1100-1400°C, or under 1 atm of flowing argon (99.999% purity) for 15 min at ≈1600°C, or in $10^{-6} - 10^{-4}$ Torr of disilane at 1200 - 1400°C for 10 to 30 min. Temperature is measured using a disappearing filament pyrometer, with calibration done by using a graphite cover over the sample and measuring its temperature.[17] A vacuum chamber connected to the graphitization chamber permits low-energy electron diffraction (LEED) measurement, using a VG Scientific rear-view LEED apparatus.

Following graphitization our samples are transferred to an Elmitec III system for low-energy electron microscopy (LEEM) and LEED measurements. Samples are initially outgassed at 700°C, and then as part of the alignment procedure in the LEEM a few ML of Pb are deposited on the sample to enable photoemission electron microscopy (since Pb has a relatively low work function). This Pb is then removed from the sample by heating it to 1050°C prior to the LEEM measurements. For Si-face graphene this heating does not produce any observable effect on the sample, but for C-face graphene the heating is found to generate a small amount of additional carbon on the surface (that is either incorporated into the graphene, or produces nano-crystalline graphite as further described in Section III). During LEEM, the sample and the electron gun are kept at a potential of − 20 kV and LEEM images are acquired with electrons having energy set by varying the bias on the sample, in the range of 0-10 eV. The intensity of the reflected electrons from different regions of the sample is measured as a function of the beam energy. These LEEM reflectivity curves shows oscillations, with the number of graphene MLs (38.0 carbon atoms/nm$^2$ in each ML) being given by the number of local minima in the curve.[21] From sequences of images acquired at energies varying by 0.1 eV, color-coded maps of the graphene thickness are generated using the method described in Ref. [4].

The vacuum system containing the LEEM is also equipped with a 5 kV electron gun and VG Scientific Clam 100 hemispherical analyzer used for Auger electron spectroscopy (AES). For routine determination of graphene thickness by AES we use the ratio of the 272 eV KLL C line to the 1619 eV KLL Si line. This ratio is analyzed with a model involving the escape depths of the electrons,[22] with the overall magnitude of the ratio being calibrated to graphene thicknesses determined by LEEM. (The model includes for the interface one ML of carbon, *i.e.* the 6√3 layer, for the Si-face, but no such layer for the C-face). LEED patterns were acquired not only in the LEEM system but also using a separate UHV system containing a VG Scientific LEED apparatus. Sample were transferred to that system through air (*i.e.* an *ex situ* measurement). The surface morphology of the graphene films was studied in air by AFM, using a Digital Instruments Nanoscope III in tapping mode. Raman spectra were measured on a Raman microscope (Renishaw, inVia) with excitation wavelengths of 514 nm. All spectra were measured using a 100× microscope objective to focus the laser excitation (10 mW) onto the samples as well as to collect the scattered light.

## III. RESULTS
### A. Si-face

Graphene formation on the Si-face of SiC in a high or ultra-high vacuum environment has been well studied by many groups and is nowadays quite well understood. A complex 6√3×6√3-R30° interface layer (denoted 6√3 for short) exists between graphene and the SiC for the Si-face.[8,9,10,11,12,13] This 6√3 layer appears to act as a template layer for graphene on the Si-face, ensuring well ordered graphene on that surface. The unit cell of graphene films on the Si-face are rotated by 30° with respect to the SiC substrate, and consecutive graphene layers assume Bernal stacking as in graphite. Despite these somewhat ideal aspects of the surface structure, graphene formed on the Si-face in vacuum is *not* so ideal, with surface pits forming naturally on the surface[11] and some variation in graphene thickness occurring over the surface.

Our results for vacuum formation of graphene on the Si-face have been previously presented in Ref. [4]. We find that, for graphene thicknesses less than or equal to 2 ML, the uniformity of the graphene is rather poor. This nonuniformity likely arises from the variation in surface topography due to steps (from unintentional miscut) and from the pits that form as a result of pinning of steps by domains of the 6√3 buffer layer.[11] The resulting nonuniformity in the film thickness appears to be an inherent property of the vacuum formation, and our results are similar to those reported by others groups.[1,11,23,24]

Annealing at elevated temperatures and/or increased times leads to greater uniformity in the surface morphology and graphene thickness, albeit with an increase in the average thickness. Examples are shown in Ref. [4], where nearly layer-by-layer growth of the graphene is found for thicknesses greater than about 2 ML (although much thicker graphene is contained in the pits that are still present on the surface[25,26]). Uniform coverage of thinner films, e.g. single ML, is very difficult to achieve by annealing in vacuum. However, it was shown by two groups that the use of an argon inert-gas environment during the annealing permitted the use of higher temperatures for an equivalent thickness of graphene, since the sublimation rate of the Si is reduced by the argon.[2,3] Higher temperature then permits a more equilibrium form of the surface structure, i.e. more uniform thickness and few, if any, of the surface pits.

In Ref. [4] we presented results for graphene on the Si-face formed under argon, demonstrating uniform ML-coverage, and in Fig. 2 we present additional results from a different sample. In this case the starting wafer has a miscut of ≈0.3°, larger than typical for our nominally on-axis wafers, leading to a significant number of step bunches forming during the graphene formation (consistent with the report of Virojanadara et al.[27]). For the annealing temperatures of Fig. 2, 1470°C, a uniform monolayer of graphene is formed between the step bunches with typically 2 ML at the bunches.

LEED patterns obtained from Si-face surfaces are shown in Fig. 3, for a surface following H-etching and for the argon-prepared graphene film of Fig. 2. In the former case the pattern consists of a 1×1 arrangement of SiC spots together with very weak (1/3,1/3) and (2/3,2/3) spots associated with a √3×√3-R30° arrangement that arises from residual oxidation of the surface.[28] For the graphitized surface there are additional 1×1 spots associated with the graphene (rotated at 30° relative to the SiC spots) together with satellite 6√3×6√3-R30° spots surrounding both the primary SiC and graphene spots arising from the underlying buffer layer and possible associated distortions of the graphene layer.[8,9,10,11,12,13]

It is interesting to compare the vacuum-prepared Si-face results discussed above with those of Bolen et al. in Ref. [29]. The background pressure in their growth system is reported to be $4\times10^{-5}$ Torr, although the composition of that gas is not specified. A number of the morphological features reported by Bolen et al. are similar to those seen in our vacuum-prepared Si-face graphene above, but one exception is the presence of prominent "fingers" in their work, identified as SiC regions not yet converted to graphene. Their samples were prepared at 1475°C. Similar finger-like morphology is reported by Ohta et al. for samples prepared in Ar at 1550°C,[30] they are observed in our own work for samples similar to those of Fig. 2 prepared in argon at ≈1600°C but with graphene coverage <1 ML (not shown), and they are also apparent in the confinement controlled sublimation (CCS) results of de Heer et al.[31] The preparation temperature of Bolen et al. is significantly higher than those for an equivalent thickness of graphene in other studies employing vacuum-preparation (1200–1300°C).[4,11,23,24] For the vacuum-prepared graphene the finger-like morphology is *not* seen, even for graphene coverages <1 ML. It should also be noted that the overall uniformity of the single-ML graphene in the Bolen et al. work is considerably better than for single-ML graphene prepared in vacuum by our method or by other groups.[1,4,11,23,24] Hence, there appears to be some difference in the growth conditions between the Bolen et al. work and the other vacuum-preparation studies. Bolen et al. report that they employ SiC-coated susceptors in their apparatus; it seems likely that those susceptors produce an unintentional Si background during graphene formation, hence leading to the higher preparation temperatures and the improved graphene morphology.

One important quantity to consider with respect to the single-ML graphene is the crystallographic grain size.[32] Although the graphene on the Si-face maintains essentially perfect rotational orientation with respect to the SiC (i.e. rotated by 30°), there still may be translational domain boundaries as well as 180° rotational boundaries in the film.[33] Studies of vacuum-prepared material by surface x-ray scattering reveal that graphene on Si-face samples has mean grain size of 40 – 100 nm,[32,34,35] which is on the same scale as (or slightly smaller than) the morphological disorder of such samples.[4] We are not aware of similar measurements for graphene prepared under argon. Nevertheless, by LEEM, domains on the several-µm length scale or larger have been observed in ML-thick graphene films,[2,3,4,27] which likely represents the grain size in those films.

As an alternative to the use of an inert gas atmosphere, the use of disilane has been proposed.[5] Results for graphene formation under a $5\times10^{-5}$ Torr environment of disilane are shown in Fig. 4. Compared to the vacuum-prepared surfaces,[4] we see that the surface morphology is somewhat smoother for the disilane-prepared film and the homogeneity of the graphene is also better. The 2-ML-thick areas of graphene in Fig. 4(b) display less of the small-scale mottling seen for vacuum preparation.[4,23] As discussed above, even for vacuum preparation, the uniformity of the graphene films improves considerably for preparation temperature above about 1320°C, although it is difficult to prepare films thinner than about 2 ML at those temperatures because of the short annealing times that would be required. With the disilane background, the annealing time can be raised without an associated increase in the Si sublimation rate. Thus, with the higher temperature, improved uniformity is achieved.

**B. C-face**

In contrast to the Si-face, graphene formation on the C-face is not so well understood. The $\mathrm{SiC}(000\bar{1})$ surface itself is not as well characterized as the (0001) surface, although 2×2 and 3×3 reconstructions of the $(000\bar{1})$ surface have been observed.[7,14,15] One STM study of those surfaces was reported, in which it was found that the 2×2 or 3×3 structure survive also at the interface between graphene and the SiC (i.e. in analogy to the 6√3 structure for the Si-face).[14] Even so, however, it is not expected that the 2×2 or 3×3 structure would act as a template for the graphene, since there is no simple coincidence between their unit cell size and that of the graphene.

Several aspects of graphene formation on the C-face are reproducibly found by many research groups to be different than those for the Si-face. In vacuum, graphene forms at a significantly lower temperature on the C-face than the Si-face.[36,37] This difference, though not well understood, likely results simply from an inherent energetic instability of the C-face compared to the Si-face, i.e. with the 6√3 layer of the latter forming a stable, low-energy surface. Also, whereas the rate of graphene formation on the Si-face drops considerably as the layers form, it seems that for the C-face that the formation rate decreases less quickly with thickness.[38] Hence, it is common to form rather thick C-face graphene films, e.g. >10 ML thick. It is found by many groups that these graphene layers in these films do not stack in the Bernal manner, but rather, considerable rotational disorder occurs.[39,40] Significantly, it was demonstrated by Hass et al. that this disorder produces a band structure even for these multi-layer films that closely resembles that of single-layer graphene.[41] Hence these multilayer films on the C-face can properly be called multilayer graphene, rather than graphite.

Aside from the structural features just described for graphene on the C-face, the reproducibility between research groups regarding growth mode or growth morphology of the graphene is somewhat limited. Inhomogeneous formation of the graphene on the C-face has been reported by several groups, although the details differ between groups. Early work of Hass et al. reported poor results for C-face graphene formation in vacuum, but much improved results using an rf-induction furnace at pressure of $3\times10^{-5}$ Torr.[34] A relatively extensive study has presented by Camara et al., in which they demonstrate that under vacuum conditions of $10^{-6}$ Torr the graphitization of 6H-SiC consists of both an extrinsic process in which an existing structural defect creates a nucleating center and an intrinsic process in which the graphitization occurs everywhere on the surface.[42,43] To control these processes they cover the SiC with a cap which reduces the Si sublimation rate and quenches the intrinsic growth process, thereby enabling the growth of graphene ribbons at step bunches due to the extrinsic process.[44]

Work of other groups revealed either islanding in the initial stages of the C-face graphene formation, or an apparent inhibition in the initial growth followed by rapid growth at temperatures above some critical temperature, with these works performed at pressures between $10^{-6}$ and $10^{-5}$ Torr.[16,45] An important factor for graphene formation on the C-face is, we believe, the cleanliness of the surface (and surrounding environment). We have previously reported the presence of a silicate layer ($Si_2O_3$) on the C-face of SiC formed under argon, i.e. due to unintentional oxidation.[7] A number of the vacuum systems used by other groups for graphene formation under vacuum have only moderate base pressures, and as further discussed in the following Section we believe that unintentional surface oxidation of the SiC (making it resistant to graphitization) is a significant factor in many of the previous reports.

The results of our studies of graphene formation on the C-face in a vacuum of $10^{-8}$ Torr (consisting mainly of H) have been presented in Refs. [6] and [7]. We find that the graphene uniformly covers the surfaces, although the distribution of thicknesses in the graphene film is significantly larger than for the Si-face, particularly for relatively thick films. We believe that this thickness variation in the film occurs both because of the relatively low temperatures used for the C-face and because of the lack of a stable low-energy interface between the graphene and the C-face SiC. Planarization of the films is inhibited, at least until temperatures of about 1250°C at which point very thick films are produced.

An additional likely consequence of the reduced temperatures used for the C-face graphene formation in vacuum is that it yields restricted grain sizes in the films. Figures 5(a) and 5(b) show AFM images of the C-face film formed at 1270°C and with 15 ML average thickness. A characteristic terraced morphology of the film is formed. Figure 5(c) shows a large-area LEED pattern of the surface, showing the characteristic circular streaks associated with a C-face graphene film. The black lines on the image indicate a 60° section of the pattern (for reference, a 60° sector with the same orientation is shown on a wide-area LEED pattern of the same sample, presented below). In Fig. 5(d) we show a selected-area LEED patterns acquired over an area with diameter of 5 μm on the surface. Many such patterns were acquired from the surface and they are all practically identical, appearing very similar to the large-area LEED pattern of Fig. 5(c). With the streaks in the LEED patterns arising from multiple orientations of graphene grains (dominated by the top graphene layer in the film),[34] the fact that the selected-area patterns are the same as the large-area patterns means that the grain size in this film is considerably smaller than 5 μm. Larger grains are achieved using a disilane environment for the graphitization, as discussed below.

Returning for a moment to the AFM image of Fig. 5(a), we see that it reveals a network of nearly parallel lines (spaced by 270 nm and running about 24° counter-clockwise from vertical).

These derive from the original atomic steps on the H-etched substrates.[6] Small amounts of nano-crystalline graphite (NCG) tends to accumulate at these steps during the graphitization procedure, and this NCG is rather immobile after it sticks to the steps so that its presence is seen even after considerable reorganization of the surface morphology as has occurred for the sample of Fig. 5. Figure 5(b) shows an expanded view of the surface morphology [acquired from an area at the center of Fig. 5(a)]. In this image we now also see a network of raised (white) lines extending between the NCG deposits on the surface. These lines separate the surface area into trapezoidal shapes with typical extent of ≈300 nm. Perhaps these areas constitute crystallographic grains of the graphene, with the raised (white) lines being boundaries between grains in the film. However, it is also possible that the raised lines arise due to strain relaxation of the graphene film as it cools following graphitization. That process produces ridges on the surface (also denoted by various authors as folds, wrinkles, ripples, pleats, or puckers) because of the different thermal expansion coefficients between the graphene and the SiC.[1,45,46,47] From our own prior work we find that such ridges have typical height of ≈10 nm, as opposed to only ≈3 nm for the features in Fig. 5(b). But nevertheless, the strain-induced ridges themselves are known to vary in size,[47] so we cannot conclusively distinguish between grain boundaries or strain-induced ridges as the origin of the raised lines in Fig. 5(b). (Prior measurements by STM yield a grain size of about 50 nm for vacuum-prepared C-face graphene,[32,48] but those samples are prepared at somewhat lower temperatures than the sample of Fig. 5).

To increase the formation temperature for graphene on the C-face, one can try the same method used for the Si-face of performing the annealing in an argon atmosphere. Unfortunately, this technique is found *not* to be successful for the C-face. Rather, we observe only the formation of isolated, thick graphene islands,[7] consistent with the work of Tedesco et al.[16] In Fig. 6 we show additional results from such samples, obtained using Raman spectroscopy as a probe of the islands. The optical images reveal the islands on the left-hand side, with a featureless region on the right. The locations at which spectra were acquired are indicated. On the islands, the spectrum is typical of few-layer graphene, with prominent G and 2D lines and a weak D peak.[17,49] Off the islands, we observe a spectrum close to that of bare SiC, although still with weak graphitic features as seen by the bottom spectrum of Fig. 6 where the D (defect) peak is just as strong as the G (graphene) peak. However, this spectrum off the island is *not* intrinsic to the Ar-annealed sample, but rather it arises from post-preparation heating of the sample in our LEEM chamber (performed prior to the Raman study), as described in Section II. The inset of Fig. 6 shows the C KLL line from AES before and after that heating, with the characteristic shoulder at 270 eV clearly revealing the graphitic carbon.

As an alternative to the graphene preparation under argon, we have also formed graphene under disilane on the C-face. Typical results are shown in Fig. 7. A relatively low pressure of disilane was used there, $1\times10^{-6}$ Torr, but even so the annealing temperature needed for graphene formation was about 100°C higher than in vacuum. A moderately thick film was formed for the sample of Fig. 7, with average thickness of 4 ML. The large range of thicknesses seen in Fig. 7(d) are similar to that found for vacuum preparation. However, one notable difference between the results of Fig. 7 compared to the vacuum-prepared film of Fig. 5 is the network of raised (white) lines prominently seen in Fig. 7(a) but that are not present (or only faintly seen on a smaller length scale) in Fig. 5. We attribute these ridges to the strain-induced features arising from the different thermal expansion coefficients between the graphene and the SiC, as discussed above. As emphasized by Hass et al., the presence of such features is an indicator of a structurally ideal graphene film.[1] An additional feature of our disilane-prepared samples is that, unlike the case for vacuum preparation, they do *not* display any NCG on their surface. Apparently the presence of the additional Si on the surface acts to provide an incorporation mechanism for that carbon.

A Raman spectrum of the graphene film of Fig. 7 is shown in Fig. 8, where it is compared to a spectrum of the vacuum-prepared film shown in Fig. 3 of Ref. [6]. The vacuum-prepared film is thicker than the disilane-prepared one, so that the graphene-derived peaks labelled D, G, and 2D are correspondingly weaker for the latter. Other than that, the spectra appear quite similar, except that the disilane-prepared sample displays a large peak at 1419 cm$^{-1}$. We do not know the precise origin of this peak, although as discussed in Ref. [17] it is seen more intensely in some substrates compared to others and it appears to be related to some sort of defect or damage in the substrate itself.

Results for C-face films prepared under a higher pressure of disilane, 5×10$^{-5}$ Torr, are shown in Fig. 9. A much thinner film was formed in this case compared to Figs. 5 or 7. The surface morphology measured by AFM, Fig. 9(a), again displays raised lines (strain-induced ridges) on the surface, indicative of the presence of a graphene film. The electron reflectivity curves C or D shown in Fig. 9(c) display a minimum near 3.5 eV, similar to that seen for vacuum-prepared films [curves A and B of Figs. 2(c) or 4(c)], but they also contain a minimum near 6.7 eV which is a new feature. This new feature is even more intensely seen (at 6.4 eV) for surface areas that do not display any simple oscillations over 2.5 – 6.0 eV, as shown by curves A and B of Fig. 9(c). These unique reflectivity curves indicate some sort of new surface structure. The fact that the minimum near 6.4 – 6.7 eV is seen most intensely in curves A and B, but also for C and D, indicates that this new surface structure is present at the graphene/SiC interface, i.e. it is the bottom-most graphene or graphene-like layer, analogous to the 6√3 layer of the Si-face. It is important to note that these unique reflectivity curves are quite different than what occurs for graphene prepared in vacuum (Ref. [7]) or with low disilane pressure (Fig. 7), demonstrating that the structure of the graphene/SiC interface has changed.

Selected-area LEED studies using an aperture size of 2 μm have been performed on the sample of Fig. 9. LEED patterns acquired over many different locations of the surface all display discrete diffraction spots located at the wavevector magnitude of the graphene lattice, i.e. similar to Fig. 5(d) but with the streaks resolved into just one or a few discrete spots. The precise locations of these spots are found to vary from point to point over the surface, indicating a grain size on the order of 2 μm or larger. This result is in contrast to the data of Fig. 5 where no variation was found in the selected-area LEED patterns, indicating grain size smaller than the aperture size (a 5 μm aperture was used in that case, but similar streaks for a vacuum-prepared film were observed with a 2 μm aperture in Ref. [7]). Use of the disilane environment is thus found to lead to much larger grains and thinner films compared to vacuum preparation.

Wide-area LEED patterns obtained from various C-face surfaces are shown in Fig. 10. For a surface following H-etching, Fig. 10(a), a 1×1 pattern is obtained, together with weak √3×√3-R30° spots arising from residual oxidation of the surface.[28] For a surface heated to 1130°C in vacuum for 15 min, such that graphene formation has just begun, Fig. 10(b), a 3×3 pattern from the SiC($000\bar{1}$) surface reconstruction[7,13,14,15] together with weak graphene streaks are seen. Figure 10(c) shows the LEED pattern for the multilayer-graphene sample of Fig. 5, displaying predominantly graphene streaks. Figure 10(d) shows the LEED pattern from the disilane-prepared sample of Fig. 9, displaying SiC 1×1 spots and graphene streaks together with a complex pattern of additional satellite spots surrounding the SiC spots. This latter pattern is not understood at present, but it possibly arises from the new interface structure found in this disilane-prepared sample.

## IV. DISCUSSION

A summary of our morphological observations for vacuum-formed graphene is given in Fig. 11. We start with a substrate having a slight unintentional miscut and that has been H-etched forming a uniform step-terrace array as shown in Fig. 11(a). For the Si-face, using temperatures near 1320°C to form about 2 ML of graphene, large step bunches occur and the graphene is found to be slightly thicker at the bunches than on the terraces,[17] as pictured in Fig. 11(b). For the C-face at the same temperature, fairly large step bunches again are found, but the graphene is much thicker (≈16 ML), as in Fig. 11(d). Such films are too thick to be probed by LEEM, but we know from previous TEM studies that such films are relatively uniform in thickness (< 20% variation).[50] For the C-face at a lower temperature of about 1170°C, Fig. 11(c), we find smaller step bunches, separated by only 1-3 μm, and a wide range in graphene thickness in the domains between bunches. The lack of strain-induced ridges in these films provides evidence of a film that is not structurally continuous over the surface.[7] In Fig. 11(c) we draw the graphene as not extending over the step edges; the diagram might be somewhat exaggerated in this regard, but based on the absence of the strain-induced ridges it seems that discontinuities do occur somewhere in the film, and the step bunches is the natural location to place them.

The islanding we see for C-face graphitization in argon is similar to that seen by Tedesco et al.[16] However, unlike us, they also see a similar islanding for samples made in vacuum. We believe this likely arises from unintentional oxidation of their surfaces, since their background pressure is only $10^{-5}$ Torr. Prakash et al. report difficulty in forming thin (ML-thick) graphene on the C-face in vacuum, with nearly no graphene forming for temperatures up to 1450°C, and then at 1475°C a film with thickness of 5 ML is observed (for constant annealing time of 10 min).[45] The background pressure is about $4\times10^{-5}$ Torr in their case, so again, we consider it likely that some oxidation of the surface occurs, inhibiting the graphene formation at the lower temperatures. The electrical properties of the few-layer graphene formed on the C-face by both of these groups are good, with mobilities >10,000 cm$^2$/Vs,[51,52] but nevertheless, the formation of thin (~ML-thick) graphene that uniformly covers the C-face surface has not been achieved. The work of Camara et al. displays rather inhomogeneous graphene formation on their C-face surfaces, although this inhomogeneity is used to advantage to form well-defined nanoribbons on the surface and producing larger areas of single-ML graphene on stepped surfaces.[43,44,53] Again, we believe that this anisotropic growth is related to the moderate vacuum conditions in their graphitization chamber. Finally, it is important to note the good thickness uniformity and the relatively high mobilities (≥15,000 cm$^2$/Vs) of the ML-thick graphene on the C-face reported by Wu et al.,[54] although the precise formation conditions for that graphene are not well understood due to the confined geometry employed for the growth.

The role of oxygen in our own observations of C-face graphene formation is definitely established by its characteristic signature in the intensity vs. voltage measurements of the √3×√3-R30° LEED patterns.[7] Nevertheless, a direct measure of the partial pressure of oxygen during our Ar-preparation procedure is lacking (such a measurement is difficult in the 1-atm Ar environment). However, we have recently obtained indirect information about the presence of oxygen in the 1-atm *H environment* during our H-etching, with analogous results expected for the Ar environment. To describe these results, we first note that the normal base pressure in our preparation system of $5\times10^{-9}$ Torr is sufficiently low to prevent any significant oxidation of both the Si-face and the C-face surfaces, as revealed by the very faint (1/3,1/3) and (2/3,2/3) spots of Figs. 3(a) and 10(a). However, over a certain period of time we used our system under conditions of reduced pumping speed (when our 150 l/s main turbo pump was not operating) provided only by the 70 l/s load-lock pump with the valve to the load-lock left open. In that case our base pressure is significantly higher, about $2\times10^{-7}$ Torr. Under these conditions the outgassing rate of the chamber walls was also significantly higher than usual; if the valve to the load-lock pump was closed (i.e. as done just prior to introducing H or Ar) then the chamber pressure rose to $2\times10^{-5}$ within a few seconds, whereas the rise for our normal operating conditions is more than an order-of-magnitude less than that. Hence, under these conditions of reduced pumping, we have higher than usual oxygen partial pressures in the 1-atm H or Ar environments.

Performing surface cleaning by H-etching under these conditions of reduced pumping is found to yield relatively intense √3×√3-R30° LEED spots on the H-etched surface (not shown). The (1/3,1/3) spots of those patterns have intensity greater than the (1,0) SiC spots, and even the (2/3,2/3) spots are clearly seen. Intensity vs. energy analysis of these spots reveals that they do indeed arise from the oxidized surface.[28] Subsequent graphene formation at 1250°C on this sort of C-face surface yields *no graphene*, even though, for a nonoxidized SiC surface (i.e. made with our usual higher pumping speed), heating at the same temperature typically yields >4 ML of graphene. Thus, the influence of an oxide layer on the surface is established once again, in agreement with our prior conclusions for the Ar-prepared surfaces.[7] Although we do not know the actual partial pressure of the oxygen in the chamber during 1-atm H or Ar procedures, we do find that, for the conditions of restricted pumping in our preparation chamber, the resulting oxygen partial pressure is sufficiently high to cause the surface oxidation during both the H and Ar procedures.

When we employ disilane for the surface cleaning rather than H, we find that the resulting LEED patterns from the surface do *not* reveal any significant √3×√3-R30° LEED spots. This is the case even when restricted pumping of the preparation chamber (i.e. prior to, and during the disilane cleaning) is employed. Of course, during the disilane cleaning (pressure of $5\times10^{-5}$ Torr) the pumping of the chamber is maintained, unlike the 1-atm H (or Ar) environments for which the pumps are valved off. Thus, it is perhaps not surprising that no significant surface oxidation occurs during the disilane cleaning. Of course, the use of disilane is a well known means of removing surface oxide (likely by the formation of volatile products such as SiO).

Disilane has a significant effect on the surface chemistry not only for the surface cleaning but also during the graphitization. For C-face (or Si-face) samples graphitized in disilane, we find a total absence of NCG on the surface. (In contrast, we find that samples cleaned in disilane but graphitized in vacuum still show very significant amounts of NCG, demonstrating that the absence of NCG found during graphitization in disilane is *not* related to the presence or absence of a surface oxide). The NCG that we commonly observe during graphitization has not been widely reported by other groups, and we believe this difference is due to the presence (oftentimes unintentional) of Si in those other growth environments. Perhaps the Si acts to combine with surface C and allow that C in incorporate into the SiC (e.g. at step edges).

In any case, for our usual vacuum conditions, we are able to form continuous, thin graphene on the C-face, although the thickness of the resulting film is still much less uniform than for the Si-face (at ≤2 ML coverage the nonuniformity is similar on the C-face and Si-face, but for thicker films the thickness uniformity of the Si-face improves whereas that for the C-face continues to be poor).[7] We attribute this inhomogeneity to kinetic limitations arising from the relatively low temperature of the formation process. Under argon, severe islanding is found, which as described above we attribute that to unintentional oxidation of the surface. Finally, to overcome that oxidation, we employed a disilane environment. In that case, thin graphene layers on the C-face surface are indeed achieved, with thickness variation significantly reduced compared to our vacuum-formed films.

## V. SUMMARY

Graphene growth on SiC{0001} surfaces was compared in three different environments: UHV, an atmosphere of argon, and a background of disilane. On the Si-face the use of an argon atmosphere or disilane background results in an improvement in morphology over vacuum prepared samples. For nominally on axis Si-face samples graphitized in argon we get monolayer domains of ≈ 15 μm size. For samples with miscuts near 0.3° we find step bunches in between the monolayer graphene. These step bunches have thicker graphene, typically 2 ML thick. The graphene formed on the Si-face is found to be relatively insensitive to the vacuum conditions (unlike the situation for the C-face), with this insensitivity perhaps arising from the $6\sqrt{3}$ reconstruction of the surface forming a stable, low-energy surface termination. For Si-face samples graphitized in disilane, while we do not get large monolayer regions as in samples graphitized in argon, there is still significant improvement in morphology over samples made in vacuum.

On the C-face graphitization occurs at a lower temperature than on the Si-face and so it is common to get thick (>10 ML) graphene films. To control this graphitization and hence achieve thinner films, we graphitized in an atmosphere of argon, identical to our procedure on the Si-face. However results on the C-face are quite different from that on the Si-face. Instead of uniform layer-by-layer growth as seen on the Si-face we observed thick islands of multilayer graphene. We attribute this islanding process to unintentional oxidation of the C-face in argon, which makes the surface resistant of graphitization (so that when the graphitization finally starts, it proceeds very quickly because of the elevated temperature). For C-face samples made in disilane we obtain films that are thinner and have a larger grain size than those made in vacuum. On certain areas of these samples we obtain unique reflectivity curves in LEEM, indicative of an interface structure that is different than what occurs for vacuum preparation of the graphene. Further studies as to the precise nature of this interface structure are in progress.

13
**ACKNOWLEDGEMENTS**

Discussions with Gong Gu are gratefully acknowledged. We thank Tian Shen and R. E. Elmquist for supplying the wafer used in the study described in Fig. 2. This work was supported by the National Science Foundation under grant DMR-0856240. Use of the Center for Nanoscale Materials at Argonne National Laboratory was supported by the U.S. Department of Energy, Office of Basic Energy Sciences under contract No. DE-AC02-06CH11357

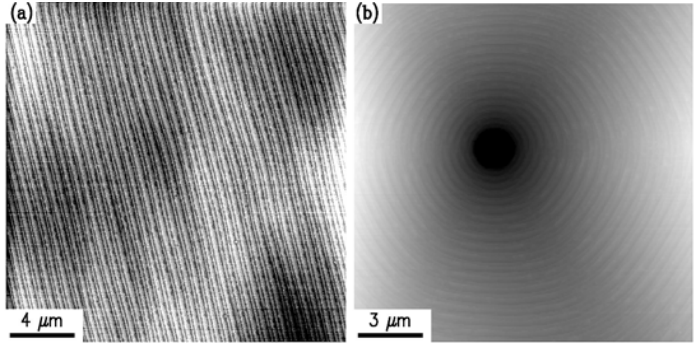

FIG 1. AFM image from C-face SiC, showing typical morphologies on (a) a surface with very few etch pits, compared to (b) a surface with many etch pits. Gray scale ranges are 2 nm and 30 nm, respectively.

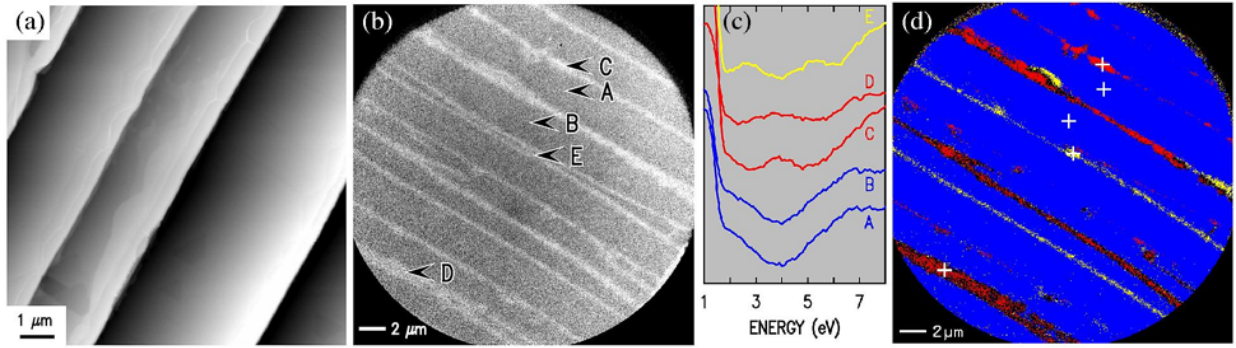

FIG 2. Results for graphene on the Si-face prepared by annealing in 1 atm of argon at 1470°C for 15 min, producing an average graphene thickness of 1.1 ML. (a) AFM image, displayed using gray scale range of 16 nm. (b) LEEM image acquired at electron beam energy of 4.4 eV. (c) Intensity of the reflected electrons from different regions marked in (b) as a function of electron beam energy. (d) Color-coded map of local graphene thickness; blue, red, and yellow correspond to 1, 2, and 3 ML of graphene, respectively. Small white or black crosses mark the locations of the reflectivity curves.

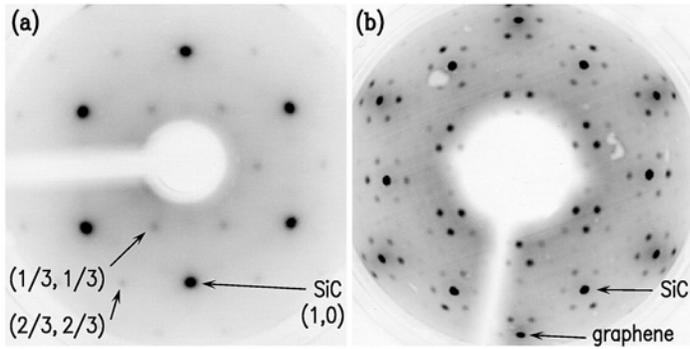

FIG 3. LEED patterns acquired at 100 eV from the Si-face: (a) following H-etching, and (b) following graphitization, for the sample of Fig. 2.

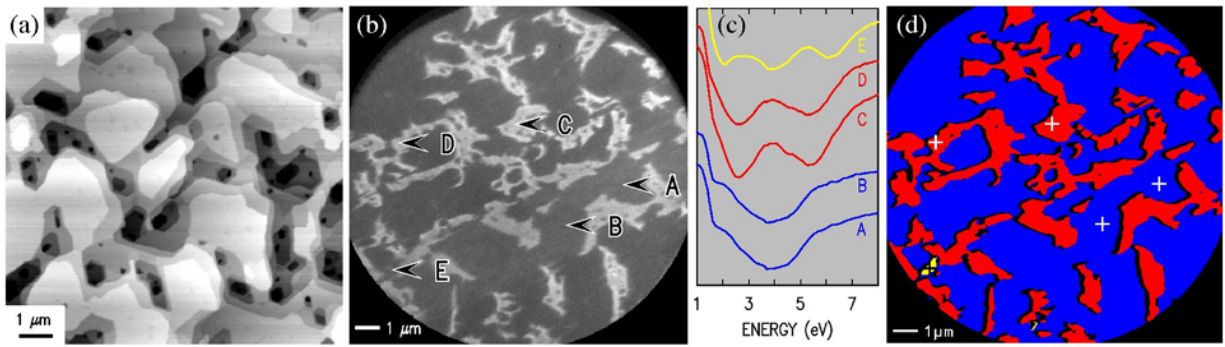

FIG 4. Results for graphene on the Si-face prepared by annealing under $5\times10^{-5}$ Torr of disilane at 1380°C for 20 min, producing an average graphene thickness of 1.3 ML. (a) AFM image, displayed using gray scale range of 3 nm. (b) LEEM image acquired at electron beam energy of 4.0 eV. (c) Intensity of the reflected electrons from different regions marked in (b) as a function of electron beam energy. (d) Color-coded map of local graphene thickness; blue, red, or yellow correspond to 1, 2, and 3 ML of graphene, respectively. Small white or black crosses mark the locations of reflectivity curves.

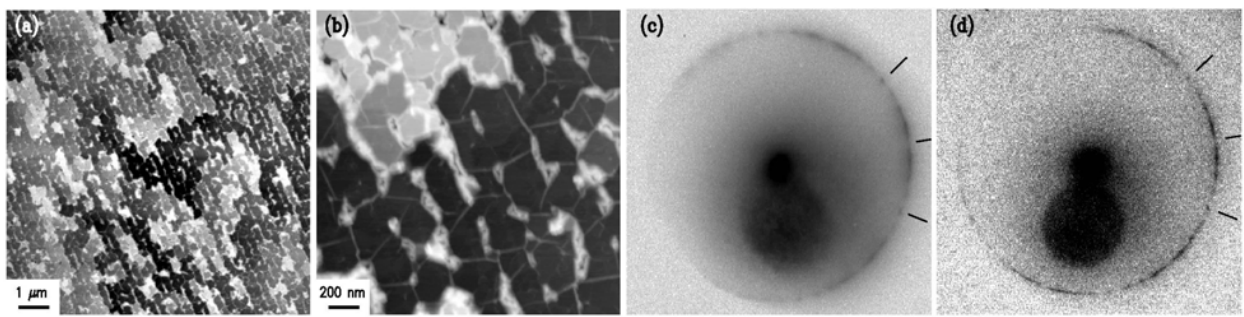

FIG 5. Results for graphene on the C-face prepared by annealing in vacuum at 1270°C for 15 min, producing an average graphene thickness of 15 ML. (a) and (b) AFM images displayed with gray scale range of 15 and 20 nm, respectively, (c) wide-area LEED pattern acquired at 44 eV over area ≈100 μm in diameter, and (d) selected-area LEED pattern acquired at 44 eV over an area ≈5 μm in diameter. Both LEED patterns reveal the characteristic circular graphene streak, with black lines indicating a 60° angular range.

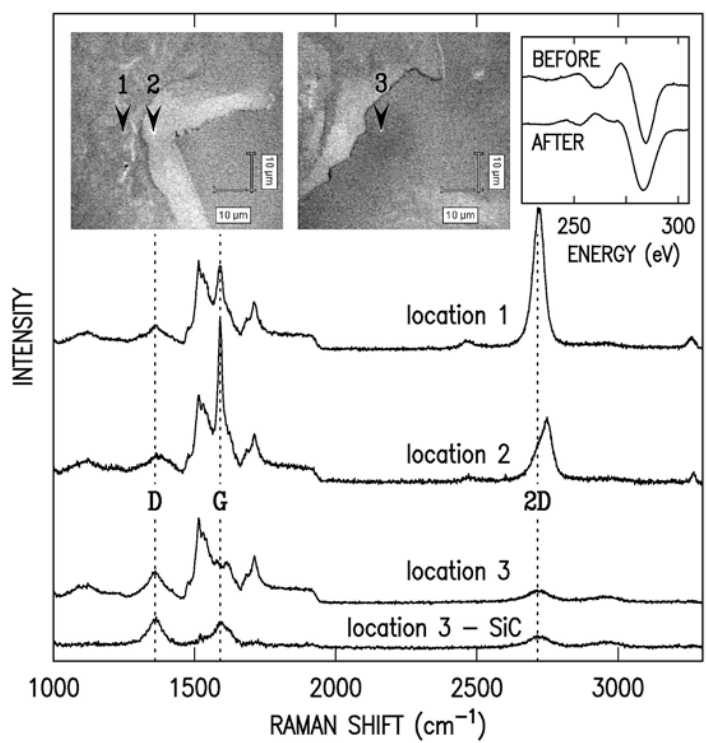

FIG 6. Raman spectroscopy results for graphene on the C-face prepared in 1 atm of argon at 1600°C for 15 min. The optical micrographs in the insets show the locations at which the spectra were acquired, locations 1 and 2 being on a graphene island and location 3 being off the island (scale bars in the insets denote 10 μm). Dotted lines are shown at the locations of the D (1360 cm$^{-1}$), G (1590 cm$^{-1}$), and 2D (2720 cm$^{-1}$) lines. Below the spectrum from location 3 is shown a difference spectrum where the bare SiC contribution has been subtracted off. The inset shows the C KLL line from AES, before and after a post-preparation heating step performed on this sample, with the characteristic graphitic shoulder at 270 eV apparent after the heating.

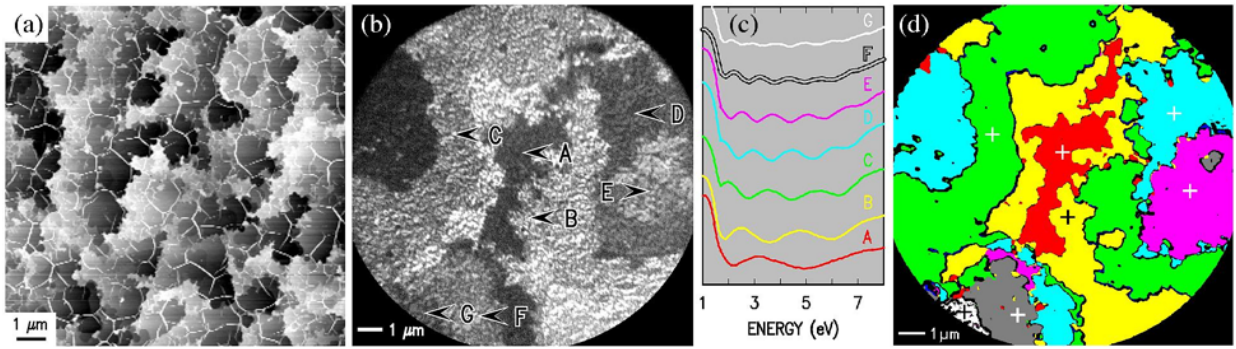

FIG 7. Results for graphene on the C-face prepared by annealing in $1\times10^{-6}$ Torr of disilane at 1290°C for 15 min, producing an average graphene thickness of 4.2 ML. (a) AFM image, displayed using gray scale range of 4 nm. (b) LEEM image acquired at electron beam energy of 2.3 eV. (c) Intensity of the reflected electrons from different regions marked in (b) as a function of electron beam energy. (d) Color-coded map of local graphene thickness; red, yellow, green, cyan, magenta, gray and white correspond to 2 – 8 ML of graphene, respectively. Small white or black crosses mark the locations of the reflectivity curves. (From Ref. [18]. Copyright Trans Tech Publications.)

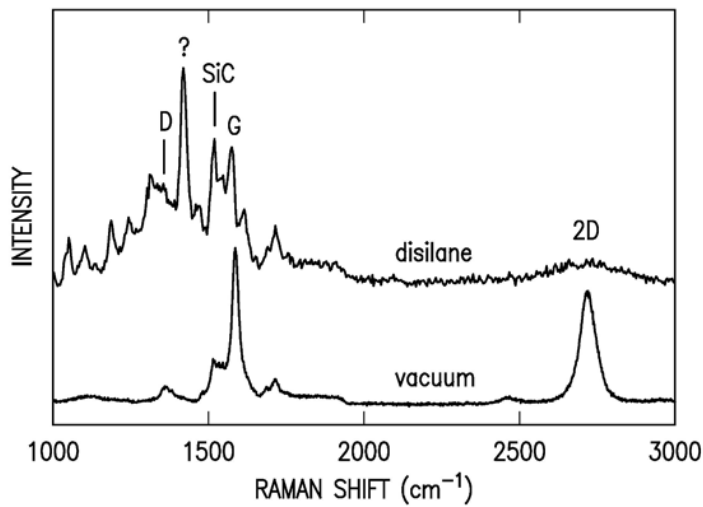

FIG 8. Raman spectroscopy from graphene on C-face SiC, obtained from the sample shown in Fig. 7 prepared in disilane (upper curve) and a sample prepared in vacuum (lower curve). The vacuum-prepared graphene is thicker, so that the graphene-derived peaks labeled D, G, and 2D are more intense for that spectrum relative to the SiC peaks (one of which is labeled). The origin of the peak marked by "?" at 1419 cm$^{-1}$ is not known.

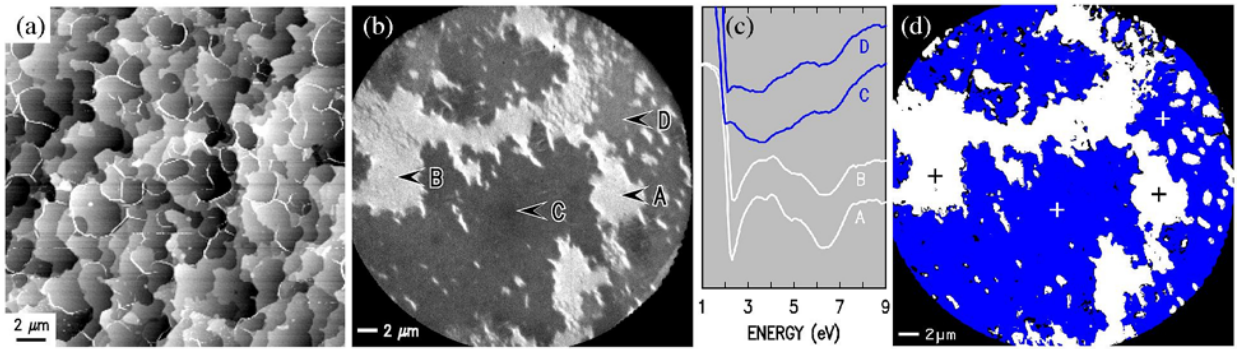

FIG 9. Results for graphene on the C-face prepared by annealing in $5\times10^{-5}$ Torr of disilane at 1270°C for 15 min, producing an average graphene thickness of 0.64 ML. (a) AFM image, displayed using gray scale range of 3 nm, respectively. (b) LEEM image acquired at electron beam energy of 4.5 eV. (c) Intensity of the reflected electrons from different regions marked in (b) as a function of electron beam energy. (d) Color-coded map of local graphene thickness; blue corresponds to 1 ML of graphene, sitting on top of an interface layer denoted by white. Small white or black crosses mark the locations of reflectivity curves.

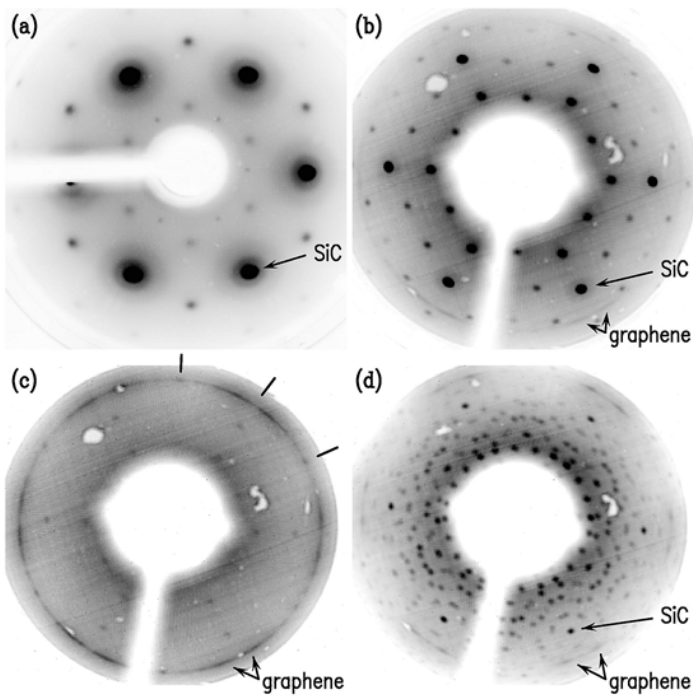

FIG 10. LEED patterns acquired at 100 eV from the C-face: (a) following H-etching, (b) following heating in vacuum to 1130°C for 15 min, (c) for the sample of Fig. 5, and (d) for the sample of Fig. 9. The black lines in (c) indicate a 60° angular range, with the same orientation as in Figs. 5(c) and 5(d).

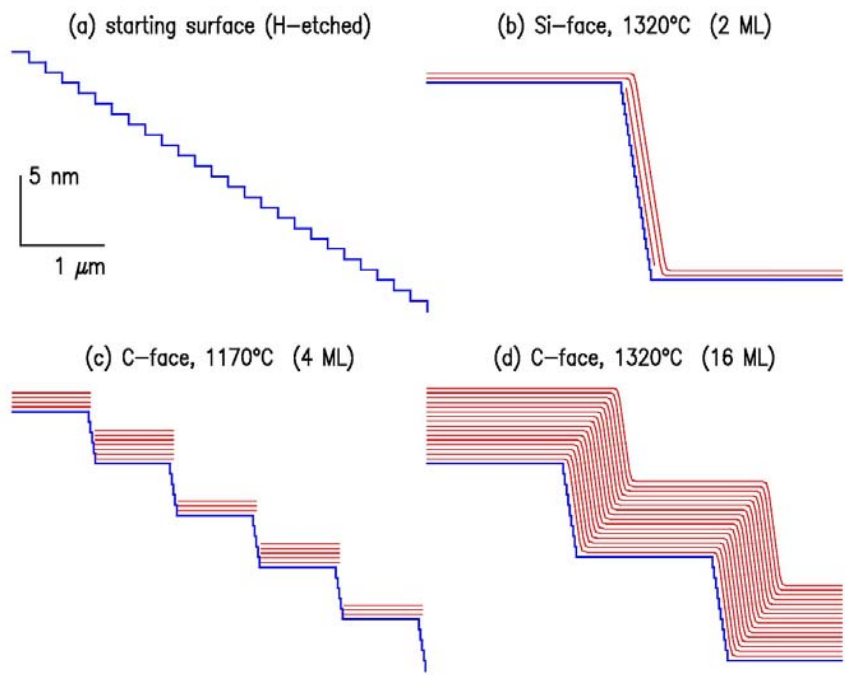

FIG 11. Schematic view of graphene formation in vacuum on Si-face and C-face SiC: (a) H-etching surface (either face). (b) Si-face after annealing at 1320°C, with large step bunches and 2 ML of graphene. (c) C-face after annealing at 1170°C, with small step bunches and 4 ML of graphene. (d) C-face after annealing at 1320°C, with large step bunches and 16 ML of graphene.